\documentclass[prd,superscriptaddress,floatfix,amsmath,footinbib,amssymb,twocolumn]{revtex4}
\usepackage{amssymb}
\usepackage{amsmath}
\usepackage{amsfonts}
\usepackage{tikz}
\usepackage{bm}

\usepackage{epsfig}
\usepackage{t1enc}
\usepackage{soul}
\usepackage{color}

\usepackage{mathrsfs}
\usepackage{url}
\usepackage[all]{xy}
\usetikzlibrary{calc,patterns,angles,quotes}

\begin{document}

\title{Reply to the comment on "Quantum priniple of relativity"}

\author{Andrzej Dragan}
\affiliation{Institute of Theoretical Physics, University of Warsaw, Pasteura 5, 02-093 Warsaw, Poland}
\affiliation{Centre for Quantum Technologies, National University of Singapore, 3 Science Drive 2, 117543 Singapore, Singapore}

\author{Artur Ekert}
\affiliation{Centre for Quantum Technologies, National University of Singapore, 3 Science Drive 2, 117543 Singapore, Singapore}
\affiliation{Mathematical Institute, University of Oxford, Woodstock Road, Oxford OX2 6GG, United Kingdom}

\maketitle

In a recent paper \cite{Dragan2020} we have investigated consequences of the Galilean principle of relativity involving inertial observers moving with {\em any} speed (not only sumbluminal one). We derived the only two mathematically allowed families of observers. In the $1+1$ dimensional spacetime case the subluminal observers are characterized by the usual Lorentz transformation:
\begin{eqnarray}
\label{Lorentz}
x' &=& \frac{x-Vt}{{\sqrt{1-V^2/c^2}}},\nonumber \\
t' &=& \frac{t-Vx/c^2}{{\sqrt{1-V^2/c^2}}}
\end{eqnarray}
valid for $V<c$ and the superluminal transformation is given by:
\begin{eqnarray}
\label{SuperLorentz}
x' &=& -\frac{V}{|V|}\frac{x-Vt}{{\sqrt{V^2/c^2-1}}},\nonumber \\
t' &=& -\frac{V}{|V|}\frac{t-Vx/c^2}{{\sqrt{V^2/c^2-1}}}
\end{eqnarray}
for $V>c$. We have shown \cite{Dragan2020} that taking into account both families of observers on equal footing, leads to three major consequences: (1) a non-deterministic decays, (2) necessity for quantum superpositions that characterize motion along multiple paths at once, and (3) complex probability amplitudes. 

In a recent paper \cite{DelSanto2022} Del Santo and Horvat question our conclusions (1) and (2) presenting three counterarguments to (1) and four counterarguments to (2). Here we show that most of their counterarguments can be dismissed, and the rest provides further evidence to our claims. 

First of all the precise formulation of the principle of relativity depends on the dimensionality of spacetime. In the simplest $1+1$ dimensional case considered here it is sufficient to state that no preferred inertial frame of reference exists and all observers witness the same set of events. One particularily important aspect of that principle is that if a certain physical process takes place in one inertial frame of reference, for example, there exist a local and deterministic mechanism ("local hidden variables") responsible for the emission of a particle at a certain moment, then such a mechanism will also exist in all other frames. 

Let us first address the three counterarguments against our claim about the inevitable indeterminism emerging from the Galilean principle of relativity.

{\bf Counterargument 1.} "One may turn the story around and claim that it is definitely the case that there is a local deterministic model in frame $O$ and that it merely seems as if there was none in frame $O'$.

{\bf Reply.} Such a scenario is only possible in superdeterministic theories, in which all physical systems are interconnected, and no two systems can be considered independent. If one excludes such theories then it is possible to consider independent objects. If two such objects exchange a superluminal particle than there is nothing in the past of the receiver that determines the moment of absorption. Consequently, the same process from the point of view of the observer for which the order of events is reversed may not be deterministic, because the moment of emission is not locally determined. Therefore the above mentioned scenario would inevitably lead to a preferred (subluminal) inertial frame of reference since it is not possible to characterize such a process using any local and deterministic theory consistently in all frames.  

On the other hand the exchange of a superluminal particle can be treated as an indeterministic process consistently in all inertial frames of reference. This is particularly clear when the sender and the receiver are identical (subluminal) particles. In this case all subluminal frames of reference should display analogous description. It would not be satisfactory that "there is a local deterministic model in frame $O$ and it seems as if there was none in frame $O'$".

{\bf Counterargument 2.} "The authors have clearly intended for the argument to apply only to microscopic phenomena."

{\bf Reply.} We have not. Although throwing of a billiard ball in the air appears to be a completely deterministic process, fundamentally such a process boils down to multiple interactions on the microscopical level, which are not deterministic. Therefore classical determinism is only approximate. To show one of the relevant aspects of the quantum-to-classical transition consider a decay of a muon, which is indeterministic on a scale of microseconds. After one second, such a decay is almost certain and therefore it is deterministic to a good approximation. Although at a certain "microscopical" time scale any decay process is indeterministic, at much larger "classical" time scales the decay probability approaches unity and the decay becomes asymptotically certain.

{\bf Counterargument 3.} "The argument put forward by the authors does not lead merely to the negation of local determinism, but to the negation of any qualitative local explanation whatsoever (be it deterministic or probabilistic)."

{\bf Reply.} An explicit example of a probabilistic theory invalidating the above claim is the one that is actually realized in nature. All fundamental decay processes have no memory. A decay rate of a muon is constant in time and does not depend on its history. Such a decay mechanism is compatible with our argument and if we were to introduce a similar emission mechanism for tachyons, it would be both probabilistic and local consistently in all frames of reference.

And now let us move on to discuss the counterarguments to our conclusion (2).

{\bf Counterargument 1.} "There are surely detectors that do not destroy their pertaining objects upon registering them. (...) One may simply construct a device that absorbs and re-emits the object in question."

{\bf Reply.} This is true and in this case our argument does not apply. Therefore one can argue that the superposition can only be witnessed if the superposed object cannot be detected without being destroyed or affected. Notice that in quantum theory non-demolishion measurements involve non-linear interactions, in which case the identity of a detected particle is ambiguous, since non-linearities are also responsible for changing the number of particles. Also once a particle is detected in one path, re-emitting it will not preserve the coherence with other paths and superpositions will inevitably be destroyed. Therefore the argument put forth by Del Santo and Horvat leads to an interesting conclusion that Galilean relativity leads to superpositions of trajectories but only for objects that cannot be detected without destroying them. Objects that can be measured without being affected are not expected to exhibit non-classical behavior, which is in perfect harmony with elementary quantum-mechanical considerations.

{\bf Counterargument 2.} "The detector placed in A-M will click with probability 1 and the second detector will never click."

{\bf Reply.} This is only true in scenarios with a preferred arrow of time. A theory without the preferred arrow of time will not have such a distinction and we only consider such theories in our work.

{\bf Counterargument 3.} "There is a whole class of quantum superposition phenomena that cannot be captured by a mere change of reference frame."

{\bf Reply.} This is true and we never claimed otherwise. Our claim is that the emergence of superpositions is inevitable, not that for any quantum-mechanical experiment there exist an inertial observer, for which the entire process appears to be purely classical.

{\bf Counterargument 4.} "One may take (the example from) the figure to represent a billiard ball bouncing from the edge of a billiard table, which would be described within a superluminal reference frame as the billiard ball travelling in superposition; however, billiard balls manifestly travel through definite trajectories."

{\bf Reply.} This is analogous to Counterargument 3. Since billiard balls can be detected (approximately) without being disturbed, one cannot argue that the superluminal observer will witness any signs of superposition.

To summarize, we reiterate our claims that taking into account superluminal observers rules out a classical, deterministic image of reality, in which objects always move along single trajectories. Local and deterministic theories are inconsistent with Galilean principle of relativity and motion along multiple paths is inevitable and can be witnessed in certain scenarios.


\begin{thebibliography}{9}

\bibitem{Dragan2020}
A. Dragan and A. Ekert, New J. Phys. {\bf 22}, 033038 (2020).

\bibitem{DelSanto2022}
F. Del Santo and S. Horvat, arXiv:2203.03661 [quant-ph] (2022).

\end{thebibliography}
\end{document}